# EVENT HORIZONS OF NUMERICAL BLACK HOLES[*]

PETER ANNINOS[1], DAVID BERNSTEIN[1], STEVEN BRANDT[1], JOSEPH LIBSON[1], JOAN MASSÓ[1,3], EDWARD SEIDEL[1], LARRY SMARR[1], WAI-MO SUEN[2], PAUL WALKER[1]

[1] *National Center for Supercomputing Applications*
*605 E. Springfield Ave., Champaign, Illinois 61820*
[2] *McDonnell Center for the Space Sciences, Department of Physics*
*Washington University, St. Louis, Missouri, 63130*
[3] *Departament de Física, Universitat de les Illes Balears,*
*E-07071 Palma de Mallorca, Spain*

ABSTRACT

We have developed a powerful and efficient method for locating the event horizon of a black hole spacetime, making possible the study of the dynamics of event horizons in numerical relativity. We describe the method and apply it to a colliding black hole spacetime.

## 1. Introduction.

Black holes are among the most fascinating predictions in the theory of General Relativity. During the past 20 years there have been intense research efforts on black holes and their effect on the astrophysical environment. The black hole events likely to be observed by the gravitational wave observatories involve highly dynamical black holes, e.g., two black holes in collision. The most powerful tool in studying such highly dynamical and intrinsically non-linear events is probably numerical treatment. In recent years, there has been significant progress in numerical relativity in this direction. In particular, long evolutions of highly dynamical black hole spacetimes are now possible,[1] opening up the opportunity of many interesting studies.

The essential characteristics of a black hole in relativity are its horizons, in particular, the apparent horizon (AH) and the event horizon (EH). Singularity theorems[2] link the formation of black hole singularities to the formation of the AH. The membrane paradigm[3] characterizes black holes by the properties of its EH, which is regarded as a 2D membrane living in a 3D space, evolving in time and endowed with many everyday physical properties like viscosity, conductivity, entropy, etc. This provides a powerful tool in obtaining insight into the numerical studies of black holes, once the horizons can be determined for numerically constructed spacetimes.

Most studies in numerical relativity on black hole horizons are in terms of the AH. The AH at each instant of time is generically a spacelike surface defined as the outermost trapped surface of a region in space, whereas the event horizon (EH) is generically a null surface defined as the boundary of the causal past of the future null infinity. The AH is a local object, readily obtained in a numerical evolution by searching a given time slice for closed surfaces whose outgoing null rays have zero expansion. In contrast, the EH is a globally defined object which makes it harder to find in numerical studies. Its location, and even its existence cannot be determined without knowledge of the complete four geometry of the spacetime. We have recently developed a powerful and efficient method to locate the EH in numerical relativity for many dynamical spacetimes of interest.[4]

---

[*] gr-qc/yymmnnn
To appear in *General Relativity* (MG7 Proceedings), R. Ruffini and M. Keiser (eds.), World Scientific, 1995.

## 2. Finding the Horizons.

There are two basic ideas in our methods in finding the horizon. They are:
(1) Integrating backward in time. We note that as going forward in time, outward going light rays emitted just outside the EH will diverge away from it, escaping to infinity, and those emitted just inside the EH will fall away from it, towards the singularity. This implies that in the time-reversed problem, any outward going photon that begins near the EH will rapidly be *attracted* to the horizon if integrated *backward* in time. In integrating backwards in time, the initial guess of the photon does not need to be very precise as it will converge to the correct trajectory after only a short time. We built our first generation of EH finder based on this backward photon method. Our second generation of EH finder has another important ingredient:
(2) Rather than independently tracking all individual photons starting on a surface, we follow the entire null surface itself, as the EH is generically a null surface. A null generator of the null surface is guaranteed to satisfy the geodesic equation. A null surface defined by $f(t, x^i) = 0$ satisfies the condition

$$g^{\mu\nu}\partial_\mu f \partial_\nu f = 0. \tag{1}$$

Hence the evolution of the surface can be obtained by a simple integration,

$$\partial_t f = \frac{-g^{ti}\partial_i f + \sqrt{(g^{ti}\partial_i f)^2 - g^{tt}g^{ij}\partial_i f \partial_j f}}{g^{tt}} \tag{2}$$

This method is simpler, more accurate and performs better in handling caustic structures of the EH than the backward photon method.

Using this method we are able to trace accurately the entire history of the EH in a short period of time. It takes just a few minutes to trace the EH on a computer workstation for an axisymmetric black hole spacetime resolved on a grid of 200 radial by 53 angular zones and evolved to $t = 75M$ (where $M$ is the mass of the black hole). We contrast our backward surface method with another method[5] that uses forward integration of individual photons to find the EH.

## 3. Horizon of two Black Hole Collision.

In Fig. (1) we show a geometric embedding of the coalescing horizons for the head-on collision of two black holes, as discussed in Ref.[1] The embedding, which preserves the proper surface areas of the horizons, shows not just the topology but also the geometric properties of the horizon. Although such a picture of the embedding is familiar, this is the first time it has actually been computed. The horizon shown in Fig. (1) corresponds an initial proper separation of the two holes of $8.92M$, where $M$ is the mass of each hole.

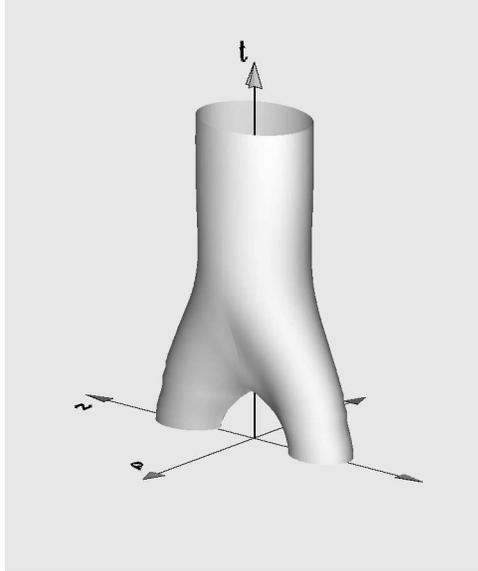

Figure 1: The geometric embedding of the event horizon for two black holes colliding head on is shown. The $z$ coordinate marks the symmetry axis, and $t$ is the coordinate time. Initially the two holes are separate, and they coalesce into a single black hole.

## 4.  Conclusions.

We have developed a powerful method for finding black hole event horizons in dynamic spacetimes based on the ideas of (*i*) backward integration and (*ii*) integrating the entire null surface. This opens up the possibility of studying the dynamics of event horizons in numerical relativity.

This research is supported by the NCSA, the Pittsburgh Supercomputing Center, and NSF grants Nos. PHY91-16682, 94-04788, 94-07882 and ASC93-18152.

## References

1. P. Anninos, D. Hobill, E. Seidel, L. Smarr, and W.-M. Suen, Phys. Rev. Lett. **71**, 2851 (1993).
2. S. W. Hawking and G. F. R. Ellis, *The Large Scale Structure of Spacetime* (Cambridge University Press, Cambridge, 1973).
3. *Black Holes: The Membrane Paradigm*, edited by K. S. Thorne, R. H. Price, and D. A. Macdonald (Yale University Press, London, 1986).
4. P. Anninos *et. al.*, Phys. Rev. Lett. (to appear).
5. S. Hughes, C. R. Keeton II, P. Walker, K. Walsh, S. L. Shapiro, and S. A. Teukolsky, Phys. Rev. D **49**, 4004 (1994).